\documentclass[aps,prl,twocolumn,superscriptaddress,nofootinbib,groupedaddress]{revtex4}
\pdfoutput=1

\usepackage{pstricks}
\usepackage{graphicx}
\usepackage{multirow}

\begin{document}

\title{Interpreting IceCube 6-year HESE data as an evidence for hundred TeV decaying Dark Matter}

\author{Marco~Chianese}
\email{chianese@na.infn.it}
\affiliation{INFN, Sezione di Napoli, Complesso Univ. Monte S. Angelo, I-80126 Napoli, Italy}
\affiliation{Dipartimento di Fisica {\it Ettore Pancini}, Universit\`a di Napoli Federico II, Complesso Univ. Monte S. Angelo, I-80126 Napoli, Italy}

\author{Gennaro~Miele}
\email{miele@na.infn.it}
\affiliation{INFN, Sezione di Napoli, Complesso Univ. Monte S. Angelo, I-80126 Napoli, Italy}
\affiliation{Dipartimento di Fisica {\it Ettore Pancini}, Universit\`a di Napoli Federico II, Complesso Univ. Monte S. Angelo, I-80126 Napoli, Italy}

\author{Stefano~Morisi} 
\email{stefano.morisi@gmail.com}
\affiliation{INFN, Sezione di Napoli, Complesso Univ. Monte S. Angelo, I-80126 Napoli, Italy}
\affiliation{Dipartimento di Fisica {\it Ettore Pancini}, Universit\`a di Napoli Federico II, Complesso Univ. Monte S. Angelo, I-80126 Napoli, Italy}

\begin{abstract}
\noindent
The assumption of a single astrophysical power-law flux to explain the IceCube 6-year HESE extraterrestrial events yields a large spectral index that is in tension with gamma-ray observations and the 6-year up-going muon neutrinos data.  Adopting a spectral index belonging to the range $\left[2.0,2.2\right]$, which is compatible with the one deduced by the analysis performed on the 6-year up-going muon neutrinos data and with $p$-$p$ astrophysical sources, the latest IceCube data show an up to $2.6\,\sigma$ excess in the number of events in the energy range 40--200~TeV. We interpret such an excess as a decaying Dark Matter signal and we perform a likelihood-ratio statistical test to compare the two-component scenario with respect to the single-component one.
\end{abstract}

\maketitle

In the last few years IceCube (IC) Neutrino Telescope has been collecting neutrino events in the TeV - PeV range~\cite{Aartsen:2013jdh,Aartsen:2014gkd,Aartsen:2014muf,Kopper:2015vzf,Aartsen:2016xlq,Aartsen:2013eka} that are providing striking evidences for extraterrestrial neutrinos even though their origin still remains unknown. According to the IC veto implementation, the IC data  are mainly divided in two samples: ``Medium Energy Starting Events'' (MESE) with an energy threshold of 1~TeV and ``High Energy Starting Events'' (HESE) for neutrino energies larger than 20~TeV. In both data sets, the neutrino interaction vertex is located inside the detector. The analysis of such events has triggered an interesting discussion in the scientific community about their possible astrophysical origin, as well as an intriguing connection with Dark Matter (DM). In particular, IC events have been related to the decay or annihilation of very heavy DM particles~\cite{Anisimov:2008gg,Feldstein:2013kka,Esmaili:2013gha,Bai:2013nga,Ema:2013nda,Esmaili:2014rma,Bhattacharya:2014vwa, Higaki:2014dwa,Rott:2014kfa,Ema:2014ufa, Murase:2015gea,Dudas:2014bca,Fong:2014bsa,Aisati:2015vma,Ko:2015nma,Dev:2016qbd,Fiorentin:2016avj,DiBari:2016guw,Zavala:2014dla,Anchordoqui:2015lqa,Dev:2016uxj,Chianese:2016smc,Borah:2017xgm,Boucenna:2015tra,Chianese:2016opp,Chianese:2016kpu,Hiroshima:2017hmy,Bhattacharya:2017jaw,ElAisati:2017ppn,Bhattacharya:2014yha,Bhattacharya:2016tma}. A pure astrophysical explanation of neutrino events has been also extensively discussed in the literature where different sources have been proposed. In particular, among such sources one can quote the stellar remnants in star-forming galaxies~\cite{Loeb:2006tw,Murase:2013rfa}, extragalactic Supernovae and Hypernovae remnants~\cite{Chakraborty:2015sta}, active galactic nuclei~\cite{Stecker:1991vm,Kalashev:2013vba,Kalashev:2014vya}, and gamma-ray bursts~\cite{Waxman:1997ti}. 

At the ICRC 2017 Conference the IceCube Collaboration has released the 6-year HESE data~\cite{proceedingIC}, which consist of a sample of 82 events (track + shower). The analysis performed on the neutrinos with a deposited energy above 60~TeV provides a best-fit power-law $E^{-\gamma}$ with spectral index\footnote{Note that a hard power-law behavior is predicted by the Waxman-Bahcall bound~\cite{Waxman:1998yy} according to the standard Fermi acceleration mechanism at shock fronts~\cite{Fermi:1949ee}.} equal to $\gamma^{\rm 6yr}=2.92^{+0.29}_{-0.33}$. Note that such a best-fit spectral index is larger than the one obtained by the 4-year HESE data ($\gamma^{\rm 4yr}=2.58\pm0.25$)~\cite{Kopper:2015vzf}. This is due to the fact that, during the last two years, IC experiment has not observed neutrinos with deposited energy larger than about 200~TeV.

According to multi-messenger analyses\footnote{In this case, it means to compare the predictions of the associated gamma-ray spectrum with the available measurements.}, one would expect a spectral index $\gamma\lesssim2.2$ in case of hadronic ($p$--$p$) astrophysical sources~\cite{Loeb:2006tw,Murase:2013rfa,Bechtol:2015uqb,Chakraborty:2016mvc}. On the other hand, photo-hadronic ($p$--$\gamma$) sources, that would produce a softer power-law neutrino flux, are strongly constrained by searches of spatial and temporal correlations with gamma-ray observations~\cite{Aartsen:2016lir,Aartsen:2017wea}. In this framework, if one takes a conservative approach, the ambiguity ($p$--$p$ {\it vs} $p$--$\gamma$) that concerns the nature of dominant neutrino sources does not allow to fix a reliable bound on the spectral index.

Nevertheless, the analysis restricted to the 6-year~up-going muon neutrinos data requires a spectral index $\gamma=2.13\pm0.13$ as best-fit~\cite{Aartsen:2016xlq}. Therefore, assuming isotropy (extragalactic sources) and an equal neutrino flavor ratio at the Earth, one would reasonably expect such a hard power-law for the whole diffuse TeV-PeV neutrino flux.

Hence, for the above arguments, large values for the spectral index like $\gamma^{\rm 4yr}$, and even more $\gamma^{\rm 6yr}$, seem to suggest that an additional component dominating at energies $E_\nu \leq 200$~TeV is required. The additional {\it second component}, superimposed to a single unbroken power-law flux, can be given, for instance, by an extra power-law~\cite{proceedingIC,Chen:2014gxa,Palladino:2016zoe,Vincent:2016nut,Palladino:2016xsy,Anchordoqui:2016ewn} (hidden cosmic-ray accelerators are viable astrophysical candidates for the second component~\cite{Kimura:2014jba,Murase:2015xka,Senno:2015tsn}) or by decay/annihilating DM particles~\cite{Chianese:2016opp,Chianese:2016kpu}.

\begin{figure}[t!]
\centering
\includegraphics[width=0.48\textwidth]{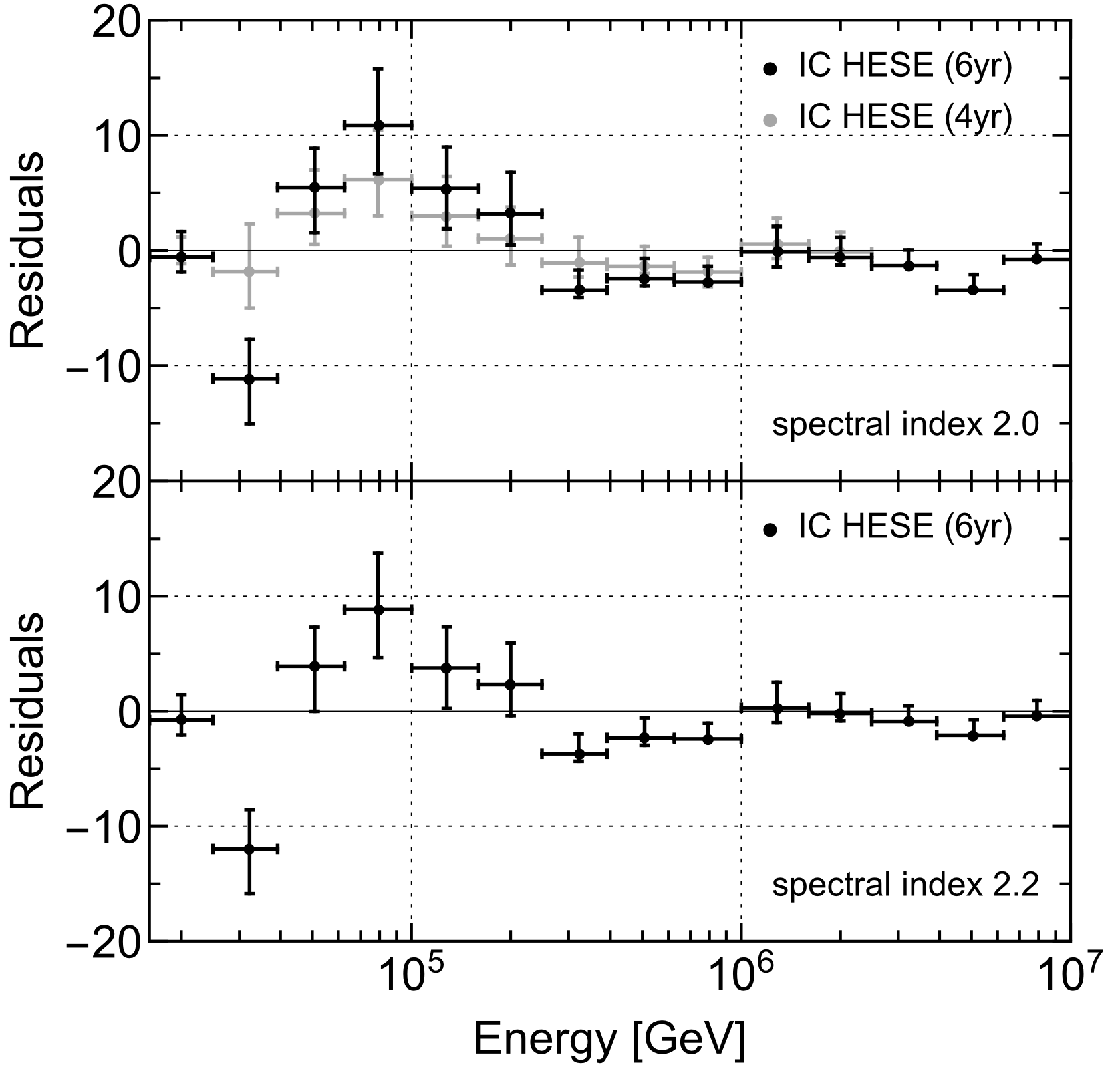}
\caption{\label{fig:residual}Residuals in the number of neutrino events as a function of the neutrino energy with respect to the sum of the conventional atmospheric background and a single astrophysical power-law with spectral index 2.0 (upper panel) and 2.2 (lower panel). We also report in gray the residuals obtained by using the 4-year HESE studied in Ref.~\cite{Chianese:2016opp}.}
\end{figure}

In Figure~\ref{fig:residual} we show the residual in number of neutrino events with respect to the sum of the conventional atmospheric background and a power-law with spectral index 2.0 (upper panel) and 2.2 (lower panel). The power-law normalization has been obtained by a maximum likelihood procedure in both cases. We highlight the presence of an excess at low energy, in the energy bins between 40 and 200~TeV. In case of spectral index 2.0, the maximum {\it local} statistical significance of such an excess is $2.6\, \sigma$ (6-year HESE), hence slightly larger than the one of $\sim2\, \sigma$ obtained in the previous analysis concerning 4-year HESE~\cite{Chianese:2016opp}. The maximum local statistical significance decreases to $2.1\, \sigma$ in case of a power-law with spectral index 2.2. We note that in case of 2-year MESE data sample~\cite{Aartsen:2014muf} the discrepancy is $2.3\, \sigma$ ($1.9\, \sigma$) for $\gamma=2.0$ ($\gamma=2.2$), as shown in Ref.~\cite{Chianese:2016kpu}. Hence, independent of the spectral index adopted, the statistical evidence for an excess slightly improves once the 6-year HESE data are used instead of  4-year HESE or 2-year MESE.

In this Letter, we focus on the interpretation of the low-energy excess in terms of two components, an {\it Astrophysical} (Astro) power-law and a decaying {\it Dark Matter} (DM) signal.\footnote{Note that the annihilation scenario is almost excluded by unitarity constraint~\cite{Chianese:2016kpu}  and by search for a signal in the Milky Way~\cite{Albert:2016emp}.} Therefore, in addition to the conventional atmospheric background~\cite{Honda:2006qj}, the whole extraterrestrial differential neutrino flux is given by
\begin{equation}
\frac{{\rm d}\phi}{{\rm d}E_\nu {\rm d}\Omega} = \frac{{\rm d}\phi^{\rm Astro}}{{\rm d}E_\nu {\rm d}\Omega} \left(\phi^{\rm Astro}_0,\gamma\right)+ \frac{{\rm d}\phi^{\rm DM}}{{\rm d}E_\nu {\rm d}\Omega} \left(\phi^{\rm DM}_0,m_{\rm DM}\right)\,.
\label{eq:tot_flux}
\end{equation}
The quantity $\phi^{\rm Astro}_0$ is the normalization of the astrophysical neutrino flux, whereas the quantity $\phi^{\rm DM}_0$ is the normalization of the DM neutrino flux, i.e. the inverse lifetime $1/\tau_{\rm DM}$ in case of decaying DM with mass $m_{\rm DM}$. For the sake of brevity we omit all the details of the above definitions. An extensive discussion of such a parametrization can be found in Ref.~\cite{Chianese:2016kpu}. Note that the {\it prompt} atmospheric background (neutrinos produced by the decays of charmed mesons)~\cite{Enberg:2008te} is here considered negligible, in agreement with the IceCube results contained in Ref.~\cite{Aartsen:2016xlq,Aartsen:2014muf,Aartsen:2013eka}.

As shown in Eq.~(\ref{eq:tot_flux}) the whole extraterrestrial neutrino flux depends on four free parameters, namely $\phi^{\rm Astro}_0$, $\gamma$, $\phi^{\rm DM}_0$ and $m_{\rm DM}$. Hence, a statistical analysis to study the relevance of the additional DM component  should consider the whole set of parameters. In this letter we follow the same approach of  Ref.~\cite{Chianese:2016kpu} that is strongly based on the prior $2.0 \leq \gamma \leq 2.2$ previously discussed. Clearly, relaxing such a constraint the evidence for an additional component becomes almost statistically irrelevant. In particular, one would obtain the best-fit $\gamma^{\rm 6yr}=2.92^{+0.29}_{-0.33}$ that is however in tension with the 6-year up-going muon neutrinos data.
In the following analysis we consider two fixed values $\gamma=2.0,2.2$ that corresponding to the extreme cases give an idea of the dependence of the results on the spectral index. To further simplify the analysis, the remaining free parameter for the astrophysical component, namely  $\phi^{\rm Astro}_0$, can be fixed at its best-fit value. This ansatz corresponds to a conservative approach where one leaves to the additional DM component the smallest room possible.

Concerning the DM component, we only consider two different decay channels of DM particles (hereafter denoted as $\chi$), which can be seen as benchmarks. In particular, we focus on the hadronic decay channel $\chi \to t\overline{t}$, and on the leptonic one $\chi\to\tau^+\tau^-$. Since the neutrino spectra produced in channels with hadronic or leptonic final-states have very different shape, considering these two decay channels practically covers all the possible phenomenological scenarios. The neutrino energy spectra provided by DM decays at the production are obtained by using the tables of Ref.~\cite{Cirelli:2010xx}. Moreover, we consider the Navarro-Frenk-White (NFW) distribution~\cite{Navarro:1995iw} for the DM halo density profile of the Milky Way and the $\Lambda$CDM parameters according to the Planck analysis~\cite{Ade:2015xua}. In the present analysis, in order to account for the neutrino oscillations we adopt the same approach  used in Ref.~\cite{Chianese:2016kpu}.
\begin{figure}[t!]
\begin{center}
\includegraphics[width=0.4\textwidth]{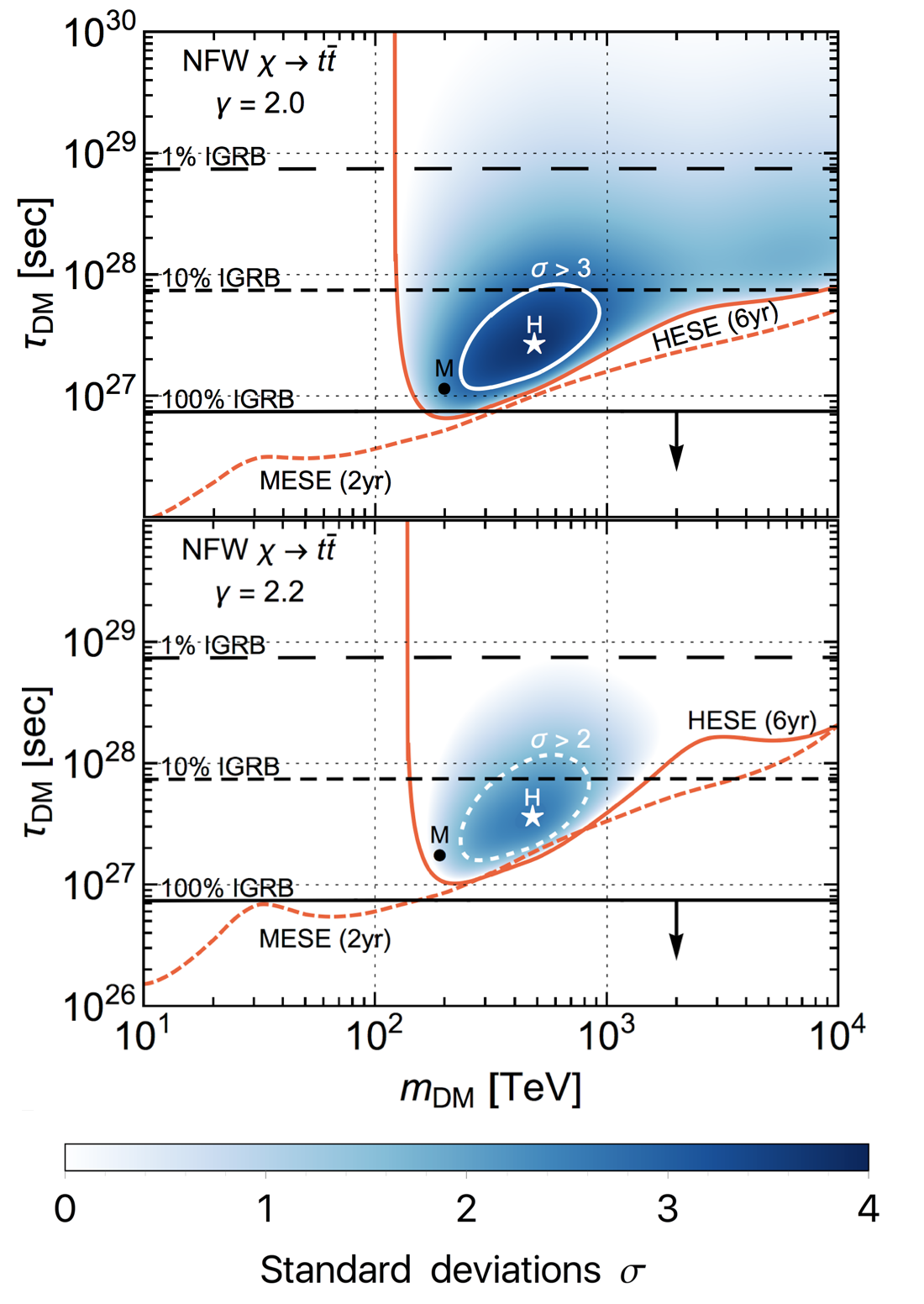}
\end{center}
\caption{\label{fig:dec_quark}Number of standard deviations $\sigma$ in the $m_{\rm DM}$--$\tau_{\rm DM}$ plane in case of decaying DM into SM quarks $\chi \to t \overline{t}$, once the spectral index of the astrophysical power-law has been fixed to 2.0 (upper panel) and 2.2 (lower panel). The white contours surround the regions where the significance of the DM component is larger than $2\, \sigma$ (dashed line) and $3\, \sigma$ (solid line). The white stars (black dots) correspond to the best-fit deduced by 6-year HESE (2-year MESE) data. The solid (dashed) red lines bound from below the allowed region according to 6-year HESE (2-year MESE) data, while the black lines represent different contributions of DM decays to the gamma-ray spectrum measured by Fermi-LAT (see the text for more details).}
\end{figure}

In order to quantify how the fit is statistically improved by adding a DM neutrino component of a given mass $m_{\rm DM}$ on top of an astrophysical power-law, we remake the analysis given in Ref.~\cite{Chianese:2016kpu} with the latest 6-year HESE data~\cite{proceedingIC}. In particular, we perform a likelihood-ratio statistical analysis on the neutrino energy spectrum, where the Test Statistics ($\rm TS$) is defined as
\begin{equation}
{\rm TS} = 2 \ln \frac{\mathcal{L}\left(\phi^{\rm Astro}_0,\gamma,m_{\rm DM}|\phi^{\rm DM}_0\neq0\right)}{\mathcal{L}\left(\phi^{\rm Astro}_0,\gamma,m_{\rm DM}|\phi^{\rm DM}_0=0\right)} \,.
\label{eq:TS}
\end{equation}
Following the above discussion, the Test Statistics is evaluated by fixing the particular DM model (final-states and $m_{\rm DM}$), the spectral index $\gamma$ and $\phi^{\rm Astro}_0$ to its best-fit value, and by scanning over the only remaining free parameter, namely the normalization of the DM flux. According to Wilks~\cite{Wilks:1938dza} and Chernoff~\cite{Chernoff} theorems, the TS follows the distribution $\frac12\delta\left({\rm TS}\right) + \frac12 \chi^2\left({\rm TS}\right)$. Therefore, we evaluate the preference of the data for a  two-component flux with respect to the single astrophysical power-law, in number of standard deviations $\sigma$. The likelihood function $\mathcal{L}$ adopted in Eq.~(\ref{eq:TS}) is a binned multi-Poisson likelihood~\cite{Baker:1983tu}, whose expression is equal to
\begin{equation}
\ln \mathcal{L} = \sum_i\left[n_i - N_i + n_i \ln\left(\frac{N_i}{n_i}\right)\right]\,,
\end{equation}
where the quantity $n_i$ is the observed number of neutrinos in the energy bin $i$, whereas $N_i$ is the expected number of events given by the sum of events related to the flux of Eq.~(\ref{eq:tot_flux}) and the background ones (conventional atmospheric neutrinos and penetrating muons). The number of extraterrestrial neutrinos is obtained by integrating the flux of Eq.~(\ref{eq:tot_flux}) with the IceCube effective area~\cite{Aartsen:2013jdh} and considering an exposure time of 2078 days (for more details see Ref.~\cite{Chianese:2016kpu}).
\begin{figure}[t!]
\begin{center}
\includegraphics[width=0.4\textwidth]{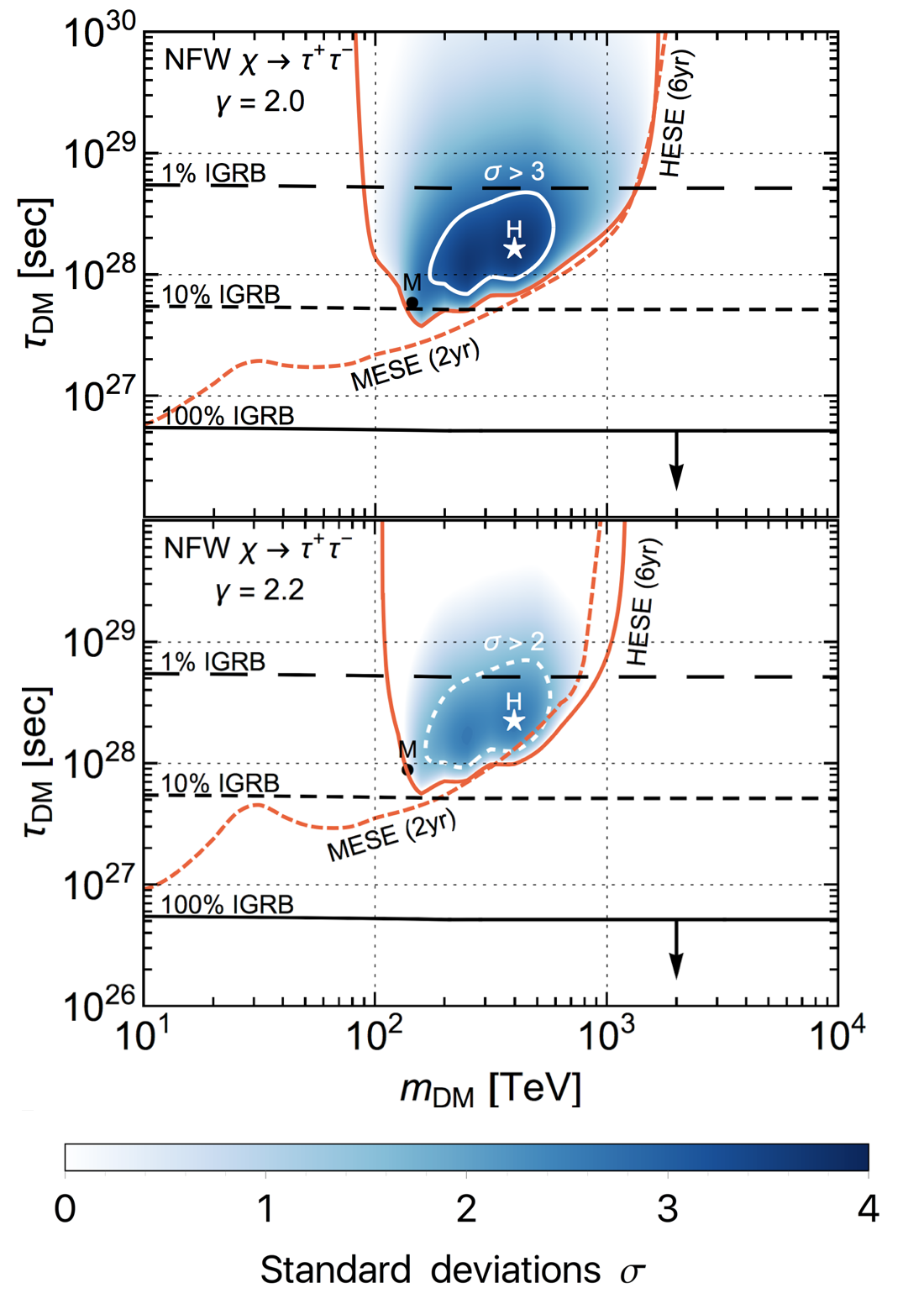}
\end{center}
\caption{\label{fig:dec_lepton}Number of standard deviations in $\sigma$ in the $m_{\rm DM}$--$\tau_{\rm DM}$ plane in case of decaying DM into SM tau leptons $\chi \to \tau^+\tau^-$. The description of the plots is the same of Figure~\ref{fig:dec_quark}.}
\end{figure}

The main results of the present analysis are presented in Figures~\ref{fig:dec_quark} and~\ref{fig:dec_lepton}. The plots display the number of standard deviations $\sigma$ in the $m_{\rm DM}$--$\tau_{\rm DM}$ plane for the decay channels considered, namely $\chi \to t \overline{t}$ and $\chi \to \tau^+\tau^-$. In particular, the darker the color, the larger the significance in $\sigma$ of the DM neutrino component. The upper and lower panels of both Figures refer to an astrophysical power-law with spectral index 2.0 and 2.2, respectively. In the plots, the best-fit values (maximum significance) is represented by white stars (the capital letter ``H'' refers to 6-year HESE analysis) and they are compared to the previous results of Ref.~\cite{Chianese:2016kpu} represented here with black dots (the capital letter ``M'' refers to 2-year MESE analysis). The white solid (dashed) contours enclose the regions in the $m_{\rm DM}$--$\tau_{\rm DM}$ plane where the statistical significance is larger than $3\,\sigma$ ($2\, \sigma$). As can be seen from the plots, the maximum value of $\sqrt{\rm TS}$ depends on the spectral index only, while it is almost independent of the decay channel considered. In particular, the statistical significance at the best-fit is $3.75\,\sigma$ and $2.60\,\sigma$ in case of spectral index 2.0 and 2.2, respectively.

Moreover, the present constraints on decaying DM models from IceCube observations are presented by the red lines. In particular, the solid red lines bound from above the regions that are excluded by the 6-year HESE data, while the dashed ones correspond to the same limit deduced by the 2-year MESE data. It is worth noting that the 6-year HESE data bound the possible DM models in a region with $m_{\rm DM}\geq 100$~TeV.  This feature depends on two effects. On one side the different energy thresholds for HESE data set (20~TeV) and MESE sample (1~TeV) provide different sensitivity of data and hence of TS for light $\chi$. On the other side as can be seen from Figure~\ref{fig:residual}, the second energy bin from left, corresponding to almost 25--40~TeV, shows a defect in the number of events and thus it disfavors any additional second component contributing to this energy. This pushes the possible DM models  to higher masses.

Furthermore, the almost horizontal black lines, instead, correspond to the gamma-ray constraints on DM models deduced by the Fermi-LAT measurements of the isotropic diffuse gamma-ray background (IGRB) spectrum~\cite{Ackermann:2014usa}. Such limits have been obtained by considering the total electromagnetic energy density~$\omega_\gamma$ of the IGRB {\it integrated} in the energy range from $0.1$~GeV to the maximum energy corresponding to $m_{\rm DM}/2$. We report the constraints for DM contribution~$\omega^{\rm DM}_\gamma$ equal to 1\%, 10\% and 100\% of~$\omega_\gamma$, respectively. In particular, the solid black lines ($\omega^{\rm DM}_\gamma = \omega^{\rm exp}_\gamma$) bound from below the allowed region in the $m_{\rm DM}$--$\tau_{\rm DM}$ plane. However, since it is quite reasonable to assume that the majority of the IGRB spectrum is accounted for by standard astrophysical sources, we consider the limit $\omega^{\rm DM}_\gamma \leq 0.1 \, \omega^{\rm exp}_\gamma$ as a realistic constraint for the DM contribution to the gamma-ray flux. Therefore, only the regions above the short dashed black lines correspond to viable choices of parameters $\left(m_{\rm DM},\tau_{\rm DM}\right)$ for DM models compatible with both neutrino and gamma-ray observations. Such multi-messenger constraints are affected by an uncertainty of about 20\%, as discussed in Ref.~\cite{Chianese:2016kpu} (see also Ref.s~\cite{Esmaili:2014rma,Murase:2012xs,Esmaili:2015xpa,Cohen:2016uyg} for different analyses about gamma-rays constraints on DM models).
\begin{figure}[t!]
\begin{center}
\includegraphics[width=0.48\textwidth]{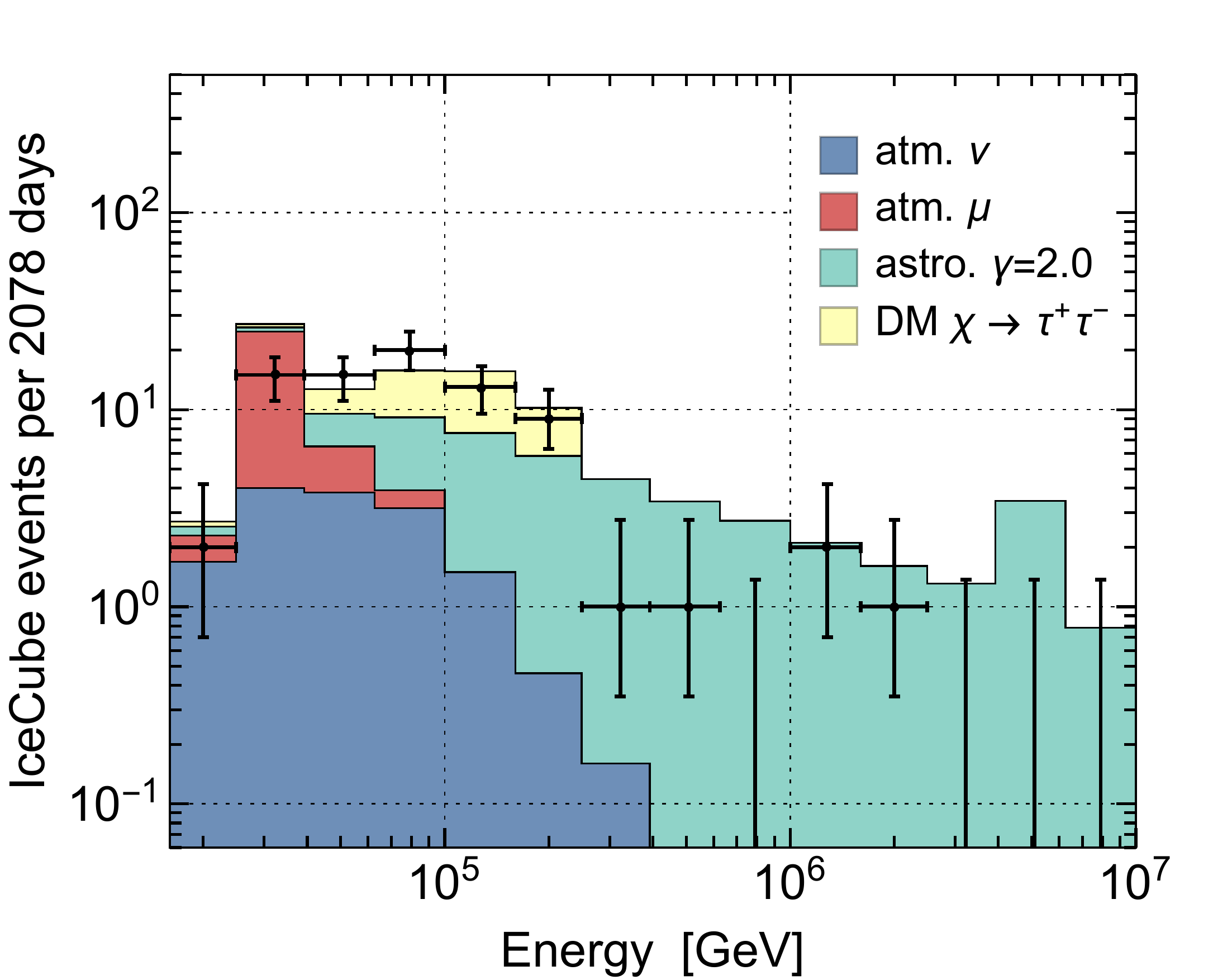}
\end{center}
\caption{\label{fig:bestfit}Numbers of neutrino events as a function of the neutrino energy after 2078 days of data-taking, for the two-component flux provided in Eq.~(\ref{eq:tot_flux}). The astrophysical contribution (green color) is a power-law with a spectral index 2.0. The DM contribution (yellow) refers to the case of decaying DM model $\chi \to \tau^+\tau^-$ with $m_{\rm DM} \simeq 400$~TeV and $\tau_{\rm DM} \simeq 1.65\times10^{28}$~sec.}
\end{figure}

We note that hadronic channel requires smaller values for the lifetime $\tau_{\rm DM}$ and larger DM masses $m_{\rm DM}$ with respect to the channel with leptons in the final-states in order to account for the IceCube data. This implies that the DM models with quarks as final-states are more in tension with Fermi-LAT data with respect to the models involving leptons, confirming the previous results discussed in Ref.~\cite{Chianese:2016kpu}. In particular, in case of a DM decaying into top quarks the best-fit values are $m_{\rm DM} \simeq 500$~TeV and $\tau_{\rm DM} \simeq 2.77\times10^{27}$~sec, while for the leptonic case we have $m_{\rm DM} \simeq 400$~TeV and $\tau_{\rm DM} \simeq 1.65\times10^{28}$~sec. In Figure~\ref{fig:bestfit} we report the predictions of the two-component neutrino flux for the latter case.

In summary, in this Letter we have focused on the latest IceCube 6-year HESE data that improve the statistics with two more years of data-taking. The IceCube analysis performed on this data sample, with the assumption of a single astrophysical power-law flux, yields a best-fit spectral index $\gamma^{\rm 6yr}=2.92$. This large value for the spectral index (even larger with respect to $\gamma^{\rm 4yr}=2.58$ of IceCube 4-year HESE) is in tension with the theoretical expectation of a hard power-law behavior for the neutrino flux ($p$-$p$ astrophysical sources) and with the analysis performed on the 6-year~up-going muon neutrinos data. Hence, it suggests the presence of a second component at low energy ($E_\nu\leq200$~TeV) added to a hard power-law flux. Here, following the same approach of Ref.~\cite{Chianese:2016kpu}, we have scrutinized the interpretation of such an additional component to diffuse neutrino flux in terms of a decaying DM signal.

The 6-year HESE data show an excess between 40 and 200~TeV, with a maximum {\it local} significance of $2.6\,\sigma$, improving the previous estimate of $2.3\,\sigma$ corresponding to 2-year MESE~\cite{Chianese:2016kpu}. The likelihood-ratio statistical test shows that the statistical improvement of the fit with the two-component flux reaches $3.75\,\sigma$ ($2.60\,\sigma$) at the best-fit for the case of $\gamma=2.0$ ($\gamma=2.2$), confirming the previous results on 2-year MESE data~\cite{Chianese:2016kpu}. However, differently from 2-year MESE, the best-fit for 6-year HESE seem to prefer larger values for $m_{\rm DM}$ even though this could be partially explained in terms of a lower sensitivity of such data set to low energy events.

\vskip5.mm

Acknowledgments. We thank E.~Cameron for the useful comments. We acknowledge support by the Istituto Nazionale di Fisica Nucleare I.S. TASP and the PRIN 2012 ``Theoretical Astroparticle Physics" of the Italian Ministero dell'Istruzione, Universit\`a e Ricerca.

\end{document}